\documentclass{PoS}
\usepackage{float}
\usepackage{graphicx,subcaption,caption}
\usepackage{amsmath,amssymb,amsfonts,bm}
\usepackage{bm}
\usepackage{xcolor}
\usepackage{lineno}
\usepackage{comment}
\graphicspath{{}}

\def \dip {D_0}
\title{Freezing out critical fluctuations}

\ShortTitle{Freezing out critical fluctuations}

\author{\speaker{Maneesha Pradeep}$^{a}$, Krishna Rajagopal$^{b}$, Misha Stephanov$^{a}$,  Yi Yin$^{c}$\\
        $^{a}$ University of Illinois at Chicago\\
        $^{b}$ Center for Theoretical Physics, Massachusetts Institute of Technology,
Cambridge, Massachusetts 02139, US\\
$^{c}$ Quark Matter Research Center, Institute of Modern Physics
, Chinese Academy of Sciences, Lanzhou, Gansu, 073000, China\\
E-mail: \email{mprade2@uic.edu}}

\abstract{ We introduce a novel freeze-out procedure connecting
  the hydrodynamic evolution of a droplet of quark-gluon plasma (QGP) that has, as it expanded and cooled, passed close to a posited critical
  point on the QCD phase diagram with the subsequent kinetic description in terms
  of observable hadrons. The procedure converts out-of-equilibrium
  critical fluctuations described by extended hydrodynamics, known as
  Hydro+, into cumulants of hadron multiplicities that can be
  subsequently measured. We introduce a critical sigma field whose
  fluctuations cause correlations between observed hadrons due to the
  couplings of the sigma field to the hadrons. We match the QGP
  fluctuations obtained via solving the Hydro+ equations describing
  the evolution of critical fluctuations before freeze-out to the
  correlations of the sigma field. In turn, these are imprinted onto
  fluctuations in the multiplicities of hadrons, most importantly
  protons, after freeze-out via a generalization of the familiar half-a-century-old
  Cooper-Frye freeze-out prescription~\cite{Cooper:1974mv} which we
  introduce~\cite{Pradeep:21}. This framework allows us to study the
  effects of critical slowing down and the consequent deviation of the
  observable predictions from equilibrium expectations
  quantitatively. We can also quantify the suppression of cumulants
  due to conservation of baryon number. We demonstrate the
  prescription in practice by freezing out the Hydro+ simulation in a
  simplified azimuthally symmetric and boost invariant background
  discussed in Ref.~\cite{Rajagopal:2019xwg}.  }

\FullConference{International Conference on Critical Point and Onset of Deconfinement\\March 15-19, 2021\\ Zoom}

\begin{document}

\section{Introduction}

A simulation of heavy-ion collisions is a multi-stage process involving
initial state dynamics, hydrodynamic evolution and a hadronic
afterburner. Transitions between subsequent stages involve matching
and translating dynamical variables used at one stage into those used
at the next stage in a way consistent with physics laws.  In this
work, we discuss the transition between the hydrodynamic stage and the
hadronic stage, for the special case of heavy-ion collisions that
produce a droplet of matter whose hydrodynamic evolution occurs in the
vicinity of a critical point on the phase diagram of
QCD. Traditionally, the macroscopic description of the quark gluon
plasma in terms of the hydrodynamic variables is translated into
a simplified hadronic description in terms of kinetic variables of an
expanding ideal resonance gas of hadrons via the well-known
Cooper-Frye procedure \cite{Cooper:1974mv}. The Cooper-Frye
freeze-out procedure has been successfully employed in the description
of high energy heavy-ion collision data for almost 50 years. The
procedure ensures that the event averaged charge and energy-momentum
densities are matched between the two descriptions.  At sufficiently
high collision energies $\sqrt{s}$, the data from many experiments are
in 
reasonable
agreement with this description across
a broad kinematic regime.

The Cooper-Frye framework, however, does not describe fluctuations.
Such a description is crucial in the special case of heavy ion
collisions that freeze out in the vicinity of a critical point. In
this case fluctuations are both enhanced and of considerable interest,
since it is via detecting critical fluctuations that we hope to
discern the presence of a critical point~\cite{Stephanov:1998dy,Stephanov:1999zu}. 
The
deviations of certain measures
of fluctuations from their non-critical baseline measured in Phase I of the
Beam Energy Scan at RHIC, deviations that vary non-monotonically
as a function of $\sqrt{s}$, 
provide intriguing hints for the possible
existence of a critical point in the QCD phase diagram
\cite{Aprahamian:2015qub,STAR:2021iop}. To convert these hints into
definitive conclusions and thereby potentially discover the location
of the critical point, one needs theoretical modeling that provides
guidance as to what to expect in this case. Much work is being done by
a number of groups to develop a description of the hydrodynamics with
critical fluctuations near a critical
point~\cite{Bluhm:2020mpc,An:2021wof}. In addition to such a
description, we need a freeze-out procedure to translate not only
average hydrodynamic variables but also the critical fluctuations
exhibited by the quark gluon plasma into the mean and fluctuations of
hadron multiplicities that are subsequently observed. This is the goal
of this work.
	 
Since the correlation length, $\xi$, grows near the critical point, the
typical relaxation time for local thermodynamic equilibration
increases. This is commonly referred to as critical slowing
down. There is a competition between the local relaxation rate and the
hydrodynamic evolution rate.  As the hydrodynamic fields evolve, if
relaxation rates are slow, as is the case for fluctuations near a
critical point, these fluctuations can no longer be described by their
equilibrium values slaved to hydrodynamic fields. There has been a
considerable amount of work performed within both the stochastic and
deterministic approaches to describe the evolution of fluctuations
near the critical point~\cite{Bluhm:2020mpc,An:2021wof}.  In the
deterministic approach, the correlation functions that describe the
fluctuations, in essence, their moments, are considered as additional
degrees of freedom with evolution equations of their own. In this
work, we introduce a prescription to convert these fluctuations, so
described, into correlations of observed particles. The results shown
here are based on work in progress~\cite{Pradeep:21}.
		
\section{Extended Cooper-Frye freeze-out near a critical point}

To describe the critical fluctuations on the kinetic side, after
freeze-out, the particle distribution function has to be modified from
an ideal gas of hadrons to one in which the correlations anticipated
near the QCD critical point are manifest. While there could be many
ways of doing this, in our work we choose to incorporate the critical
fluctuations via the interaction of the particle fields with a
critical $\sigma$ field that controls the masses of hadrons.
The model that we describe here has been discussed in equilibrium
settings and also in non-equilibrium settings in the past
\cite{Stephanov:2009ra,Athanasiou:2010kw}. In this work, we first
match the correlations of the $\sigma$ field to hydrodynamic
fluctuations.  The masses of particles at freeze-out then depend on
the background field $\sigma$. We denote $g_A$ as the coupling between
the $\sigma$ field and the field corresponding to the particle species
$A$. Small deviations of $\sigma$ from its equilibrium value change
the mass of particle $A$, to leading order as
	\begin{eqnarray}\label{massmod}
	\delta m_A=g_A \sigma
	\end{eqnarray}
and the modified distribution function is
\begin{eqnarray}\label{dismod}
f_A=\left<f_A\right>+g_A\frac{\partial \left<f_A\right>}{\partial m_A}\sigma \ .
\end{eqnarray}
In Eq.~(\ref{dismod}), $\left<f_A\right>$ is the particle distribution
function for an ideal gas. We ignore quantum statistics and viscous
corrections and use the Boltzmann distribution function
$\left<f_A\right>=\exp\left\{-\frac{p\cdot
    u}{T} + \frac{\mu_A}{T}\right\}$. We define $\sigma$ so that
$\left<\sigma\right>\equiv 0$. Hence, the mean multiplicity of particle species $A$ denoted by $\left<N_A\right>$ is left unchanged from the Cooper-Frye prescription~\cite{Cooper:1974mv}: 
\begin{eqnarray}
\label{mean1}
\left<N_A\right>&=&d_A\int dS_\mu\, \int Dp_A \,p^{\mu}\,\left<f_A(x,p)\right>\ .
\end{eqnarray}
where $d_A$ is the (iso)spin degeneracy of the particle $A$. In Eq.~(\ref{mean1}), $dS_\mu$ is a differential element on the freeze-out hypersurface pointing along the normal and $Dp_A$ is given by
 \begin{equation}
	\label{Dp1}
	Dp_A=2\frac{d^4p}{(2\pi)^{3}}\delta \left(p^{2}-m_A^2\right)\theta(p^{0}) \ .
	\end{equation}
 
 The two-point correlation function  $\left<\sigma(x_+)\sigma(x_-)\right>$ is proportional to the two-point correlation function of the slowest mode, namely fluctuations of entropy per baryon, denoted by $\hat{s}\equiv s/n$:
\begin{eqnarray} \label{ss2}
\left<\sigma(x_1)\sigma(x_2)\right>&=&Z(x)\,\left<\delta\hat{s}(x_1)\delta\hat{s}(x_2)\right>\ .
\end{eqnarray}
Here, $Z$ is a function of thermodynamic fields which depends on the QCD equation of state near the critical point.  In this work, we focus on matching the leading singular contribution to the two-point correlation function between the hydrodynamic and kinetic descriptions. With Eq.~(\ref{ss2}), the critical contribution to the variance of the multiplicity of particles of species $A$ is then given by
\begin{eqnarray}
\label{variance2}
\left<\delta N_A^2\right>_\sigma&=&g_A^2 \,\int dS_{\mu,+}\, \int dS_{\nu,-}\, Z(x)\, J_A^{\mu}(x_+)\,J_A^{\nu}(x_-)\,\left<\delta\hat{s}(x_+)\delta\hat{s}(x_-)\right>\ ,
\end{eqnarray}
where\begin{eqnarray}
	\label{JA1}
	J_A^{\alpha}(x_\pm)
	&\equiv&\frac{d_A m_{A}}{T(x_\pm)}\int\, Dp_A \,\frac{\partial \left<f_A(x,p)\right>}{\partial m_A}\, p^{\alpha}\ .
   \end{eqnarray}
     The total variance of the multiplicity of $A$ particles is the Poisson value plus the critical part, namely
   $\left<\delta N_A^2\right>=\left<N_A\right>+\left<\delta N_A^2\right>_\sigma$. 
   
\section{Demonstration in an azimuthally symmetric boost invariant
  Hydro+  simulation}

The Hydro+ framework combining hydrodynamics with a deterministic
description of out-of-equilibrium critical fluctuations was introduced
in Ref.~\cite{Stephanov:2017ghc}.  We shall use the
prescription introduced in the previous Section to freeze out a Hydro+
simulation discussed in Ref.~\cite{Rajagopal:2019xwg}. The correlation
function of $\hat{s}\equiv s/n$ can be expressed in terms of its
Wigner transform:
	 \begin{eqnarray}
\phi_{\bm{Q}}(x)=\int_{\Delta x} \left<\delta \hat
           s\left(x_+\right)\delta \hat s\left(x_-\right)\right>\,  e^{iQ\cdot\Delta x}\ .
\end{eqnarray}
Here $x=(x_++x_-)/2$ and $\Delta x=x_+-x_-$ and the integral is
performed over an equal-time hypersurface in the local rest frame at
$x$. The relaxation of this quantity to its local equilibrium value
$\bar{\phi}_{\bm{Q}}$ is governed by the equation
\cite{Stephanov:2017ghc}:
\begin{eqnarray}
	\label{phiev1}
	u(x)\cdot \partial \phi_{\bm{Q}}(x)=-\Gamma(\bm Q)\, \left(\phi_{\bm{Q}}(x)-\bar{\phi}_{\bm{Q}}(x)\right) \ ,
	\end{eqnarray}		  
where $\bar{\phi}_\mathbf{Q}$ can be adequately approximated by the
Ornstein-Zernike ansatz as in Ref.~\cite{Rajagopal:2019xwg}:
\begin{eqnarray}
	\label{phieq1}
	\bar{\phi}_{\bm{Q}}\approx\frac{c_M\xi^2}{1+(Q\xi)^2}\ .
	\end{eqnarray}
	The leading behavior of the $Q$-dependent relaxation rate $\Gamma$ near a critical point with Model H relaxation dynamics is given by:
	\begin{eqnarray}
	\label{GammaH}
\Gamma(\bm Q)=\frac{2\dip\xi_0}{\xi^{3}} K(Q\xi)\ ,
\end{eqnarray}
where $K(x)=\frac{3}{4}\left[1+x^2+(x^3-x^{-1})\arctan
  x\right]$.  Because $\phi_{\bm Q}$ describes fluctuations of the diffusive mode
$\hat s$, the rate vanishes at $\bm Q=0$, i.e., $\Gamma\approx 2D
Q^2$ where $D=\dip \xi_0/\xi$ is the diffusion coefficient, also vanishing at the critical point.

We evolve the $\phi_{\bm{Q}}$'s according to Eq.~(\ref{phiev1}) in an
azimuthally symmetric and boost invariant background using the code
described in Ref.~\cite{Rajagopal:2019xwg}. As the trajectory on the
phase diagram describing the history of a given fluid cell passes by the
critical point, the equilibrium correlation length $\xi$ increases to
a maximum value $\xi_{\rm max}$ before falling back to its value at
freezeout. We shall use variable $\xi_{\rm max}$ to describe the
effect of varying the $\sqrt s$ of the collision on how close the
trajectory passes to the critical point.  We use the parametrization
for the correlation length $\xi$ as a function of the decreasing
temperature $T$ along the trajectory from
Ref.~\cite{Rajagopal:2019xwg}:
\begin{eqnarray}\label{xiparam1}
\left(\frac{\xi}{\xi_0}\right)^{-4}=\tanh^{2}\left(\frac{T-T_c}{\Delta T}\right)\left[1-\left(\frac{\xi_{\text{max}}}{\xi_0}\right)^{-4}\right]+\left(\frac{\xi_{\text{max}}}{\xi_0}\right)^{-4}\ .
\end{eqnarray}
 In Eq.~(\ref{xiparam1}), $T_c$ is the (crossover) temperature at the point along the system's trajectory that is closest to the critical point and $\Delta T$ sets the size of the critical region. $\xi_0$ is the non-critical correlation length away from the critical point.  We set the parameters in Eq.~(\ref{xiparam1})  to  $\xi_0=0.5\, \text{fm}$, $T_c=0.16\, \text{GeV}$, and $\Delta T=0.2 \, T_c$.		
		
			\begin{figure}[t]
\begin{center}
\begin{subfigure}{0.245\textwidth}
\includegraphics[scale=0.295]{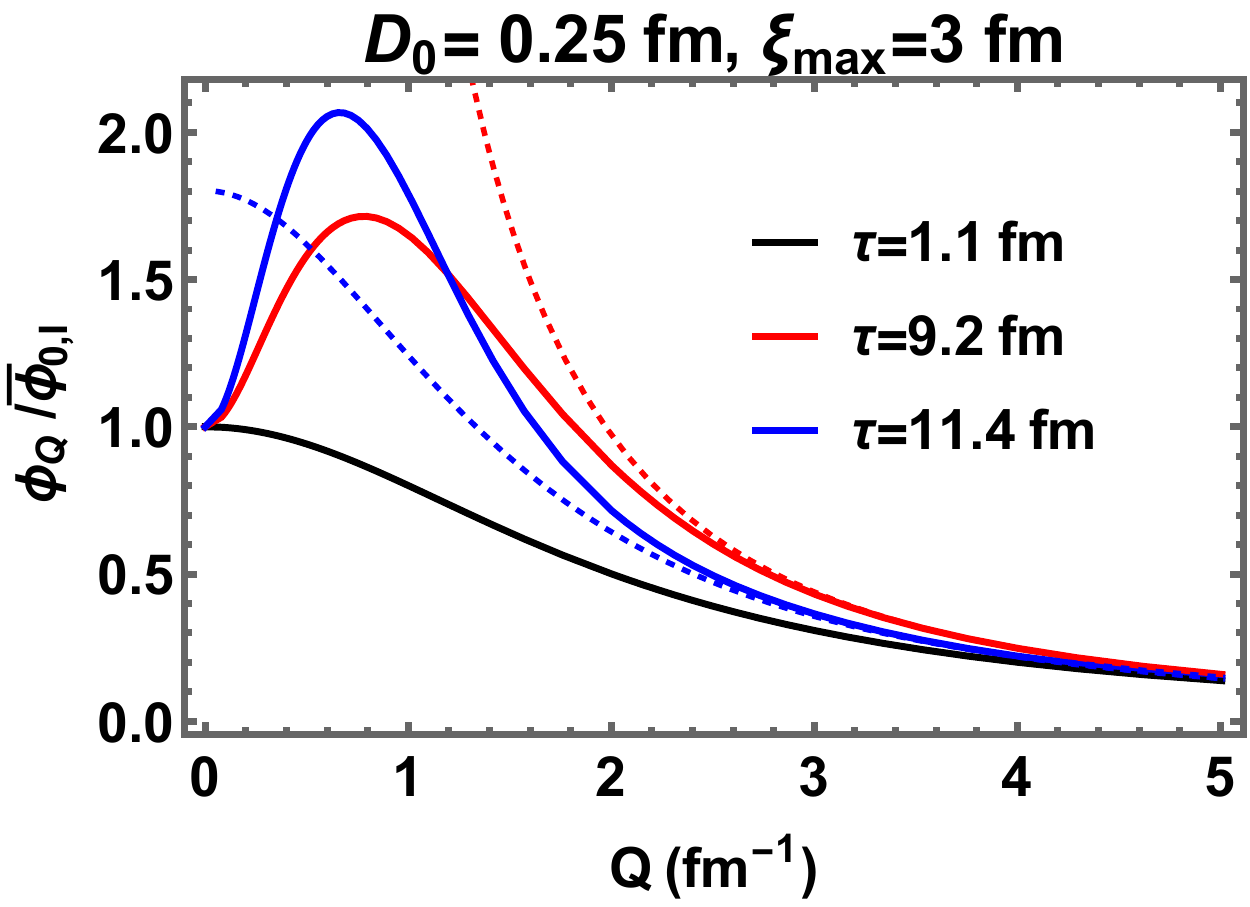}
\end{subfigure}
\begin{subfigure}{0.245\textwidth}
\includegraphics[scale=0.295]{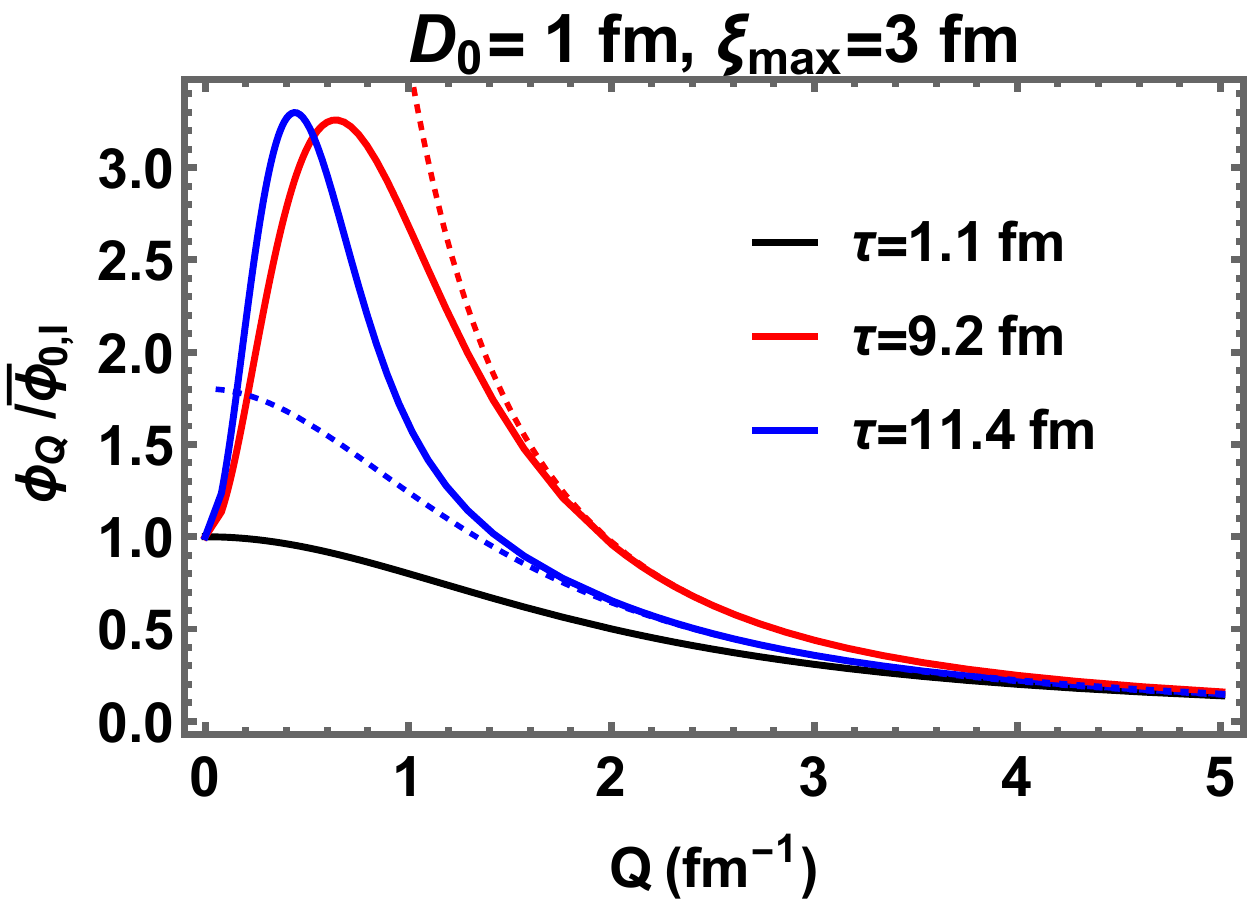}
\end{subfigure}
\begin{subfigure}{0.245\textwidth}
\includegraphics[scale=0.295]{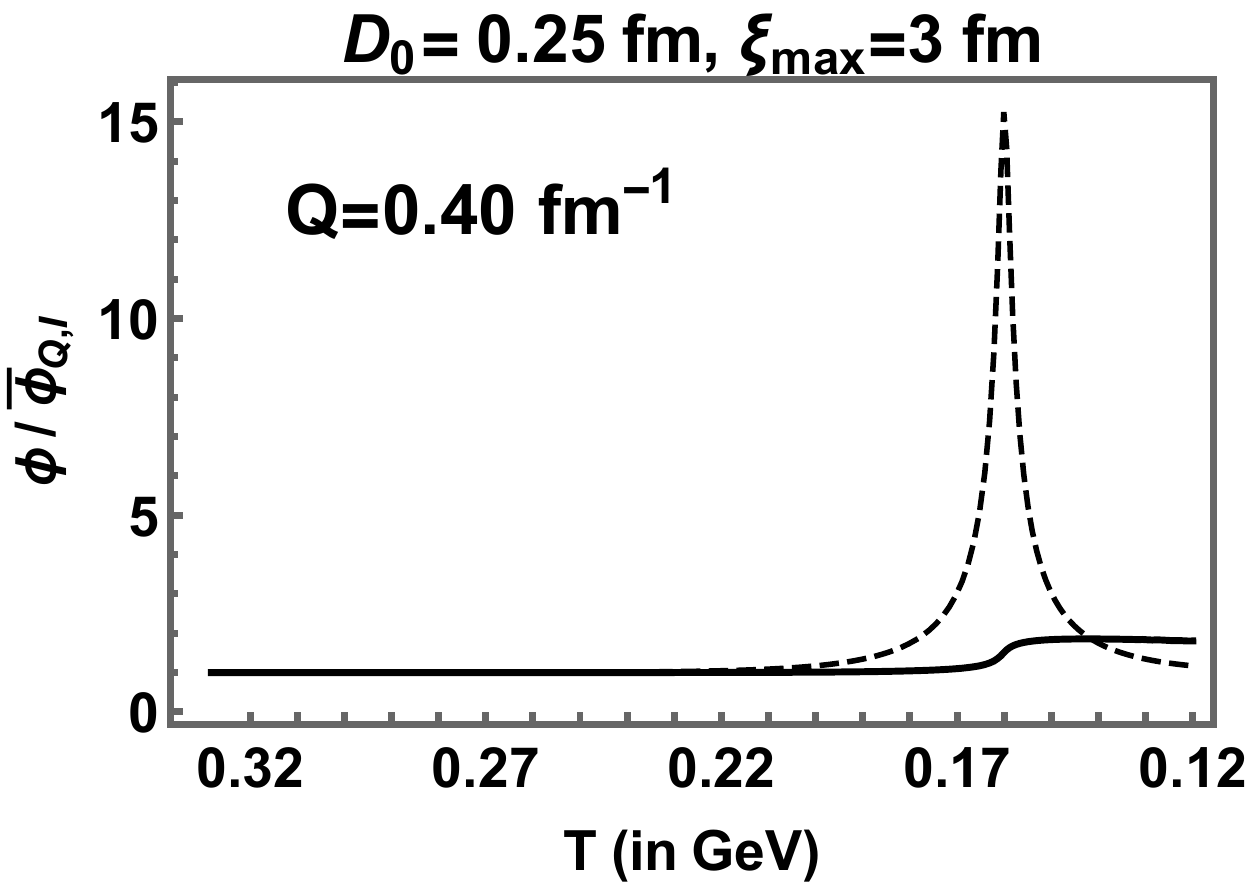}
\end{subfigure}
\begin{subfigure}{0.245\textwidth}
\includegraphics[scale=0.295]{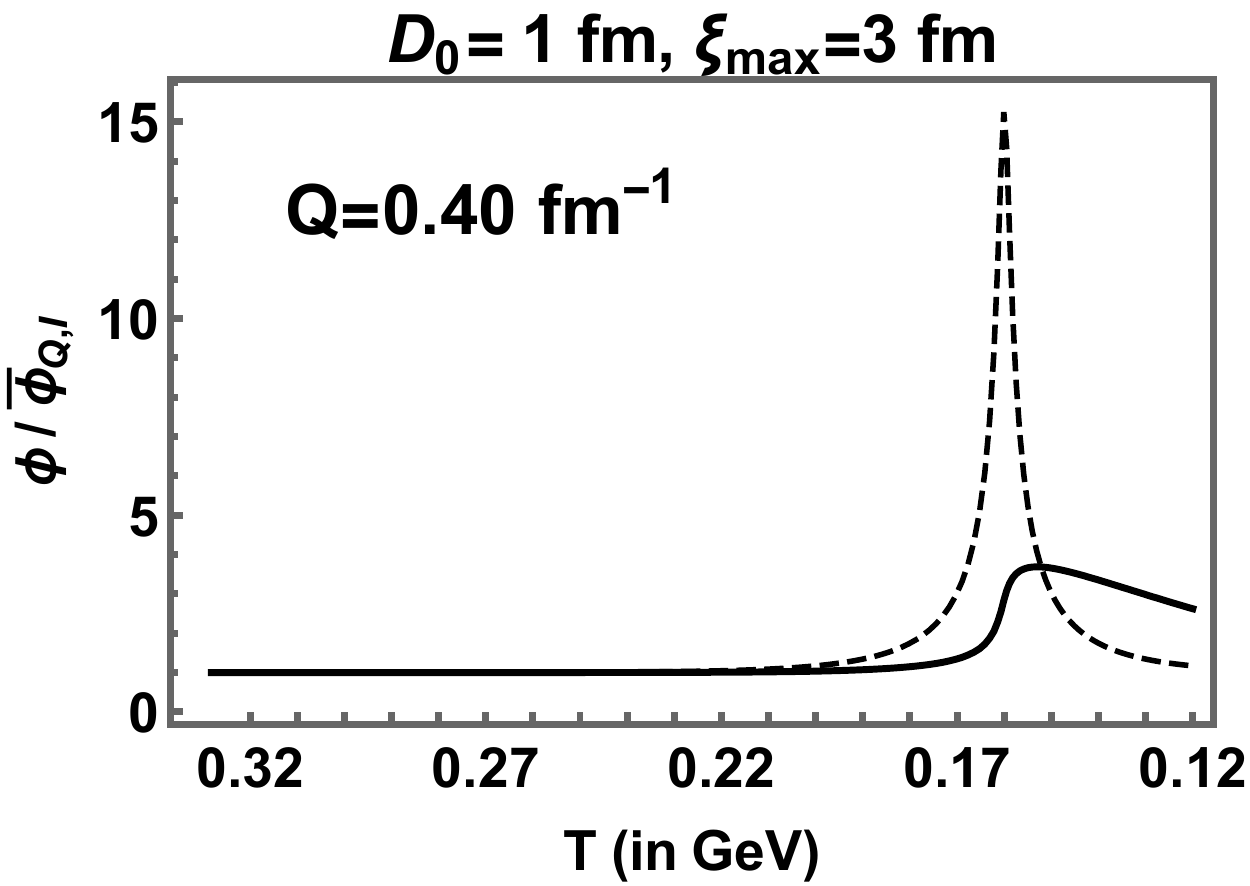}
\end{subfigure}
\end{center}
\caption{Normalized $\phi_{\bm Q}$'s as a function of $Q\equiv |\mathbf{Q}|$
  in a fluid cell near the center of the fireball are shown in the
  \textit{first} and \textit{second} plots from the left for $\dip$
  values of $0.25\, \text{fm}$ and $1\, \text{fm}$ respectively. The
  thick and dashed black curves in the these plots are the normalized
  $\phi_{\bm Q}$ and $\bar{\phi}_{\bm Q}$ at the corresponding proper
  time $\tau$ values, where $\bar\phi_{\bm Q}$ denotes the value that
  $\phi_{\bm Q}$ would have if it were in equilibrium. The
  normalization factor
  $\bar{\phi}_{\bm Q}\equiv\bar{\phi}_{\bm Q} (T(\tau_I))$.  The temperatures
  at different times are $T(9.2 \mbox{ fm})=0.14 \mbox{ GeV}$ and $T(11.4 \mbox{ fm})=0.16 \mbox{ GeV}$. The
  \textit{third} and \textit{fourth} plots from the left show the
  normalized $\phi_{\bm Q}$ and $\bar{\phi}_{\bm Q}$ as functions of the 
  temperature (i.e.,~of time) for $Q=0.4\, \text{fm}^{-1}$.}
\label{phi-1ch}
\end{figure}

\begin{figure}[t]
\begin{center}
\begin{subfigure}{0.245\textwidth}
\includegraphics[scale=0.295]{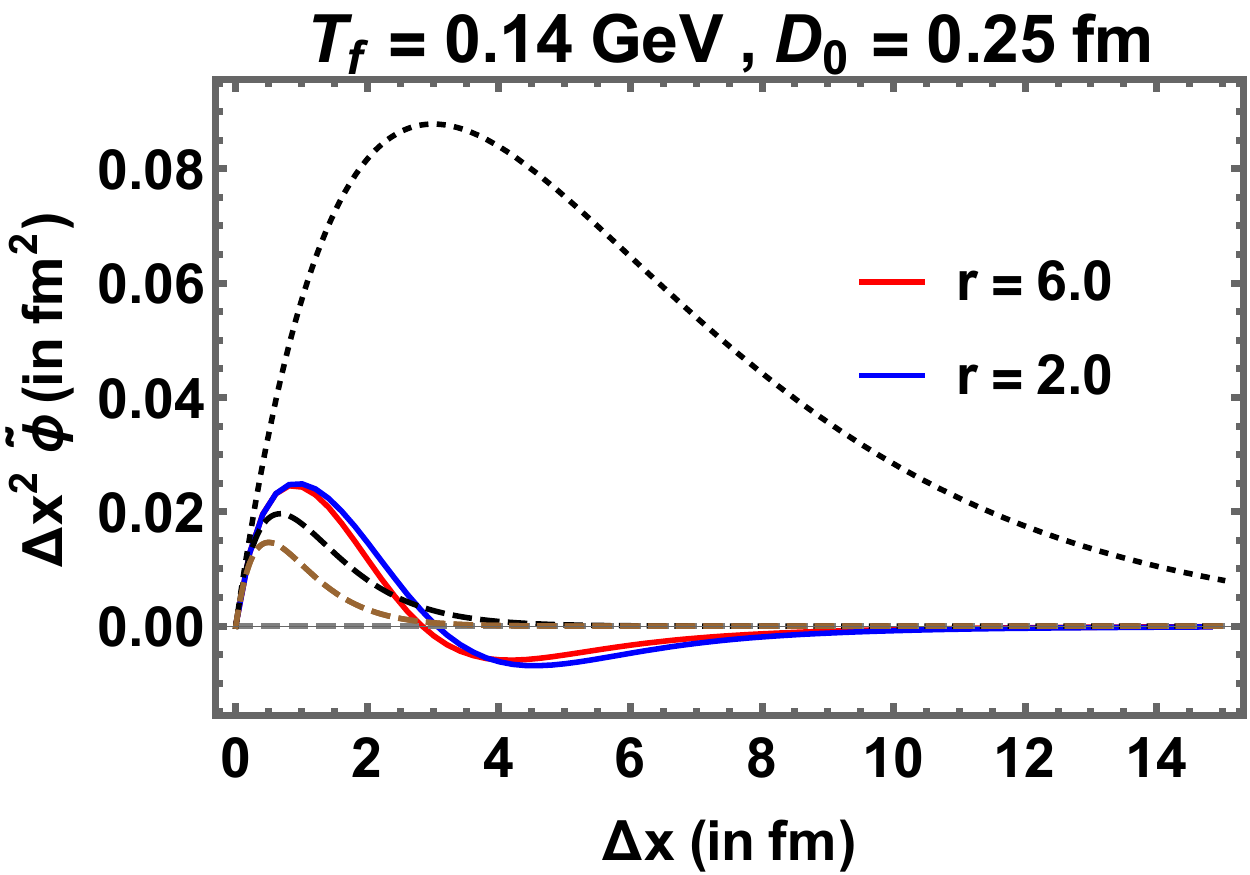}
\end{subfigure}
\begin{subfigure}{0.245\textwidth}
\includegraphics[scale=0.295]{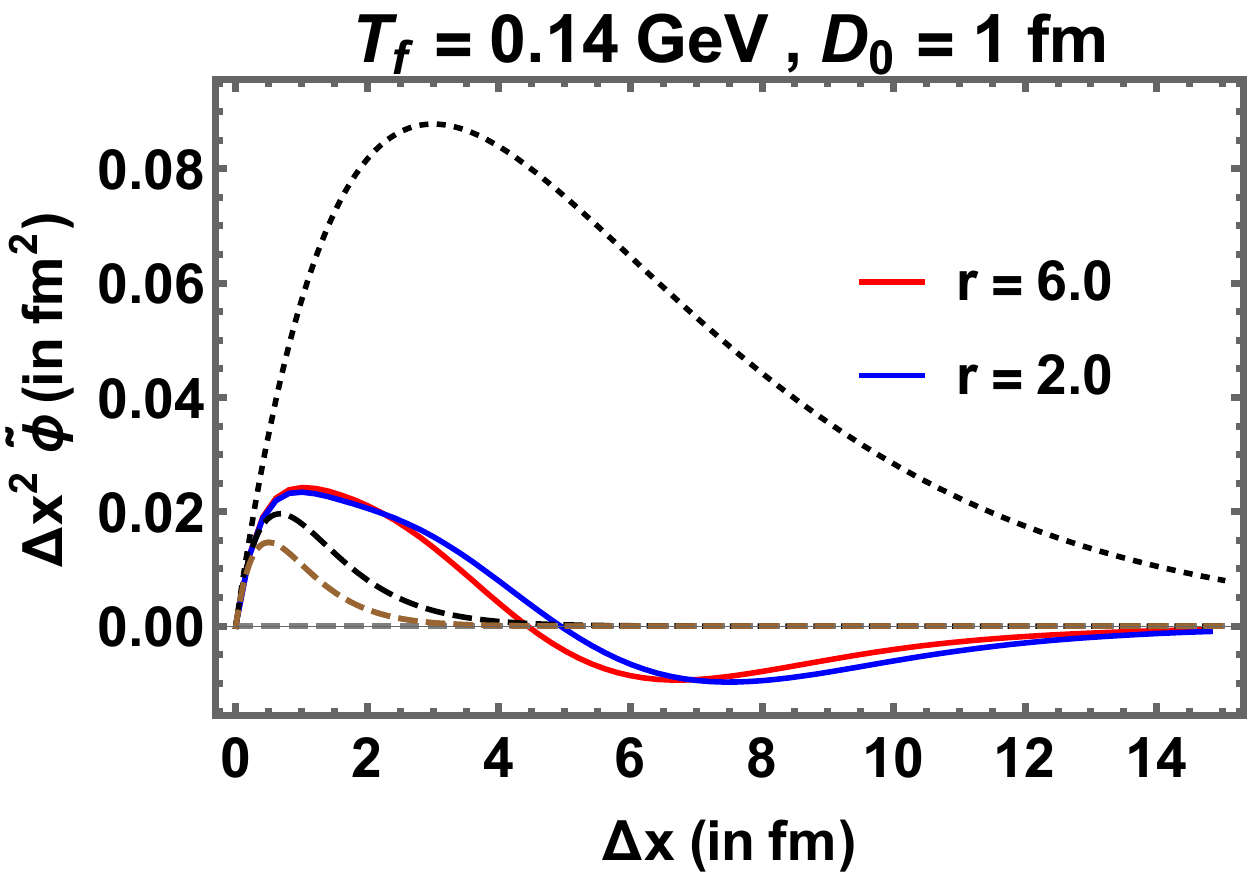}
\end{subfigure}
\begin{subfigure}{0.245\textwidth}
\includegraphics[scale=0.295]{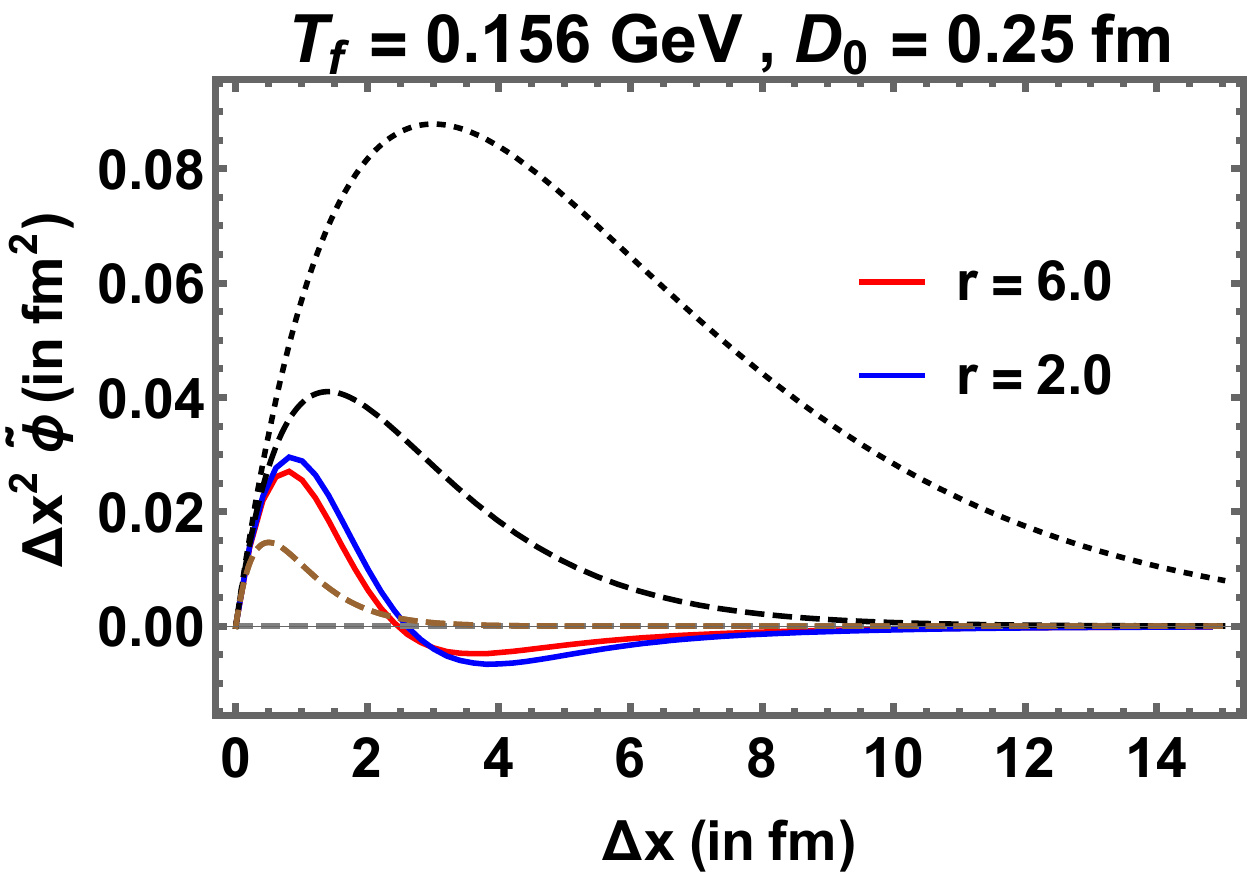}
\end{subfigure}
\begin{subfigure}{0.245\textwidth}
\includegraphics[scale=0.295]{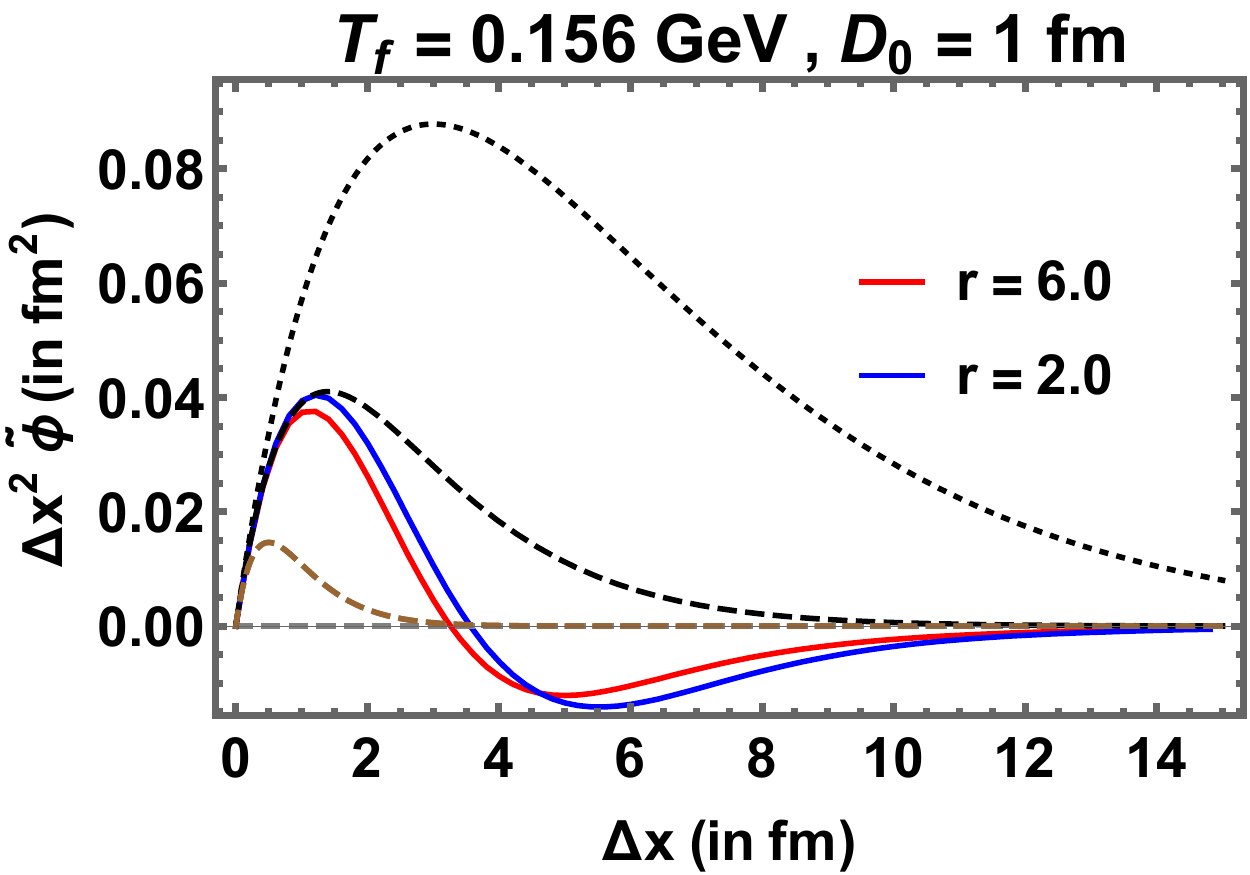}
\end{subfigure}
\end{center}
\caption{Position-space correlator of $s/n$ fluctuations $\tilde\phi$: Here, we plot
  $\Delta x^2\tilde\phi$, as a function of the spatial separation
  $\Delta x$ between the points. We plot results on two different
  isothermal freeze-out hypersurfaces  ($T=0.14$ and $0.156$~GeV) for
  two values of $\dip$'s at two radial coordinates.   The dashed black
  curve is the equilibrium correlator at freeze-out. The dashed brown
  curve is the equilibrium correlator far from the critical point,
  i.e., when $\xi=\xi_0$. The dotted black curve is the equilibrium
  correlator at $T=T_c$.}
\label{phiFT1}
\end{figure}

The $\phi_\mathbf{Q}$'s at various points along the trajectory of a
fluid cell near the center of the fireball (radial position
$r=0.7\, \text{fm}$ at initial proper time $\tau_I=1.1\, \text{fm}$) are shown
in Fig.~\ref{phi-1ch} for two relaxation rates (\ref{GammaH})
characterized by differing values of the parameter $\dip$ which specifies the diffusion constant.  The
curves in Fig.~\ref{phi-1ch} do not evolve at $\bm{Q}=0$ because  
the conservation of baryon number dictates the diffusive nature
of the $\hat s$ mode, which is a function of $n$, and its fluctuations $\phi_{\bm Q}$.
Upon increasing the $\dip$ value
from 0.25 fm to 1 fm, we see an enhancement in fluctuations as they
relax faster to their large equilibrium values near the critical point. In
this work, we considered two isothermal freeze-out scenarios namely
$T_f=0.14\, \text{GeV}$ and $T_f=0.156\, \text{GeV}$. The inverse
Wigner transform, i.e.,~the coordinate space correlation function 
$\tilde{\phi}(\Delta x, x)\equiv
\left<\delta\hat{s}(x_+)\delta\hat{s}(x_-)\right>$, is plotted for
$\dip$ equal to 0.25 fm and 1 fm on each of these freezeout surfaces in
Fig.~\ref{phiFT1}.

	\begin{figure}[t]
\begin{center}
\begin{subfigure}{0.245\textwidth}
\includegraphics[scale=0.295]{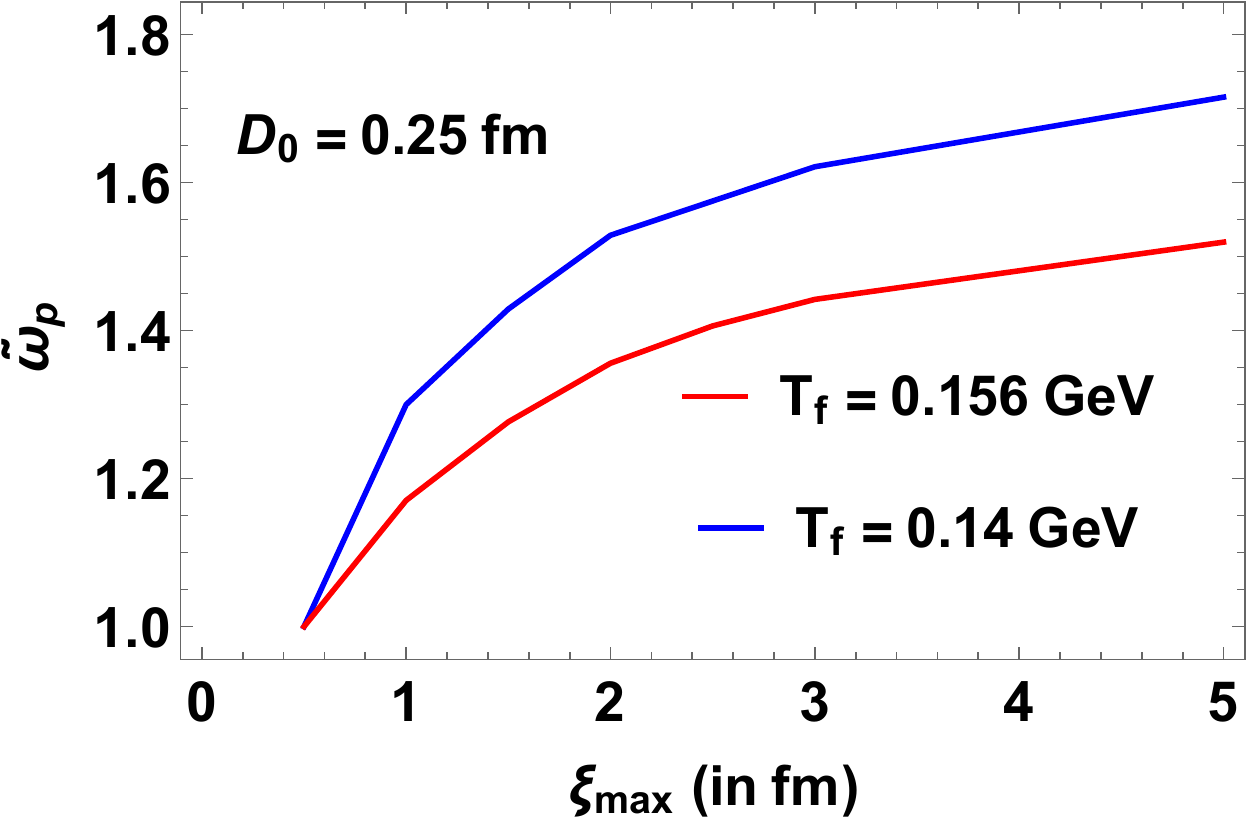}
\end{subfigure}
\begin{subfigure}{0.245\textwidth}
\includegraphics[scale=0.295]{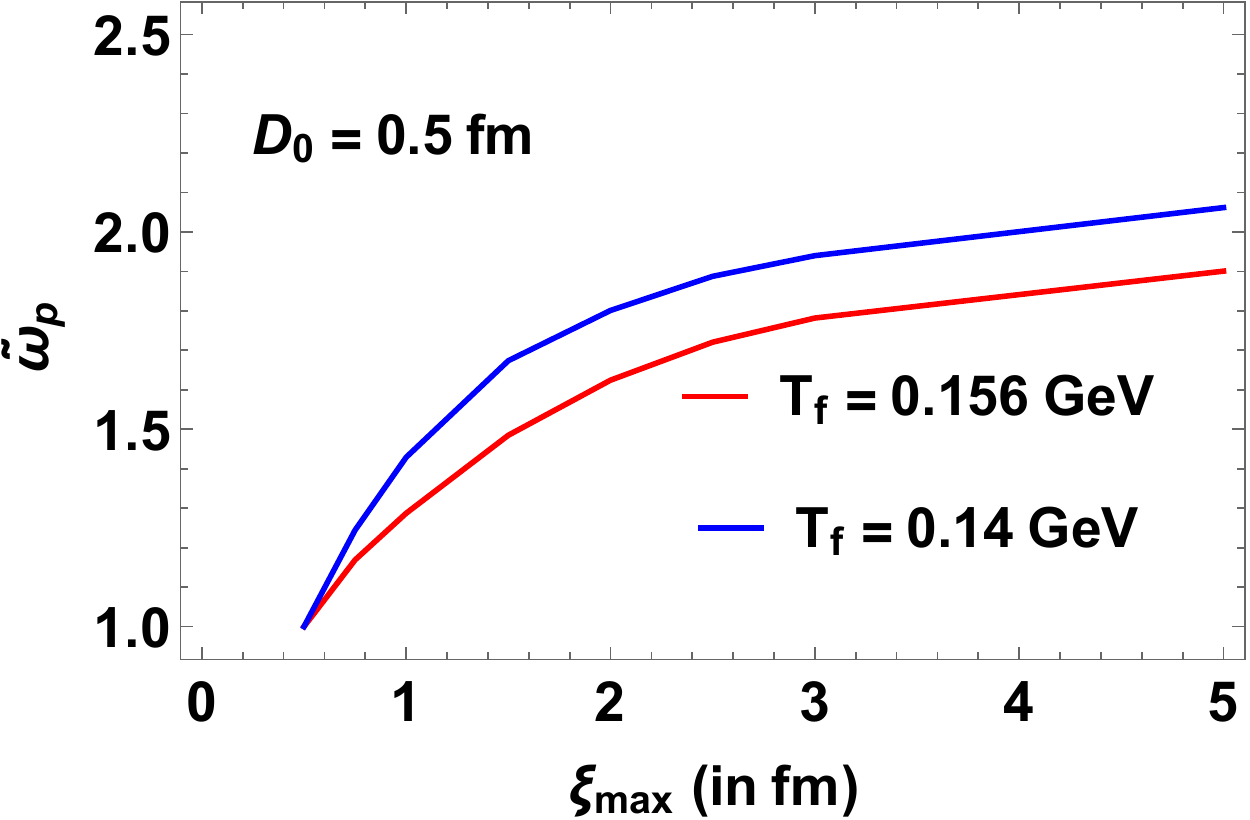}
\end{subfigure}
\begin{subfigure}{0.245\textwidth}
\includegraphics[scale=0.295]{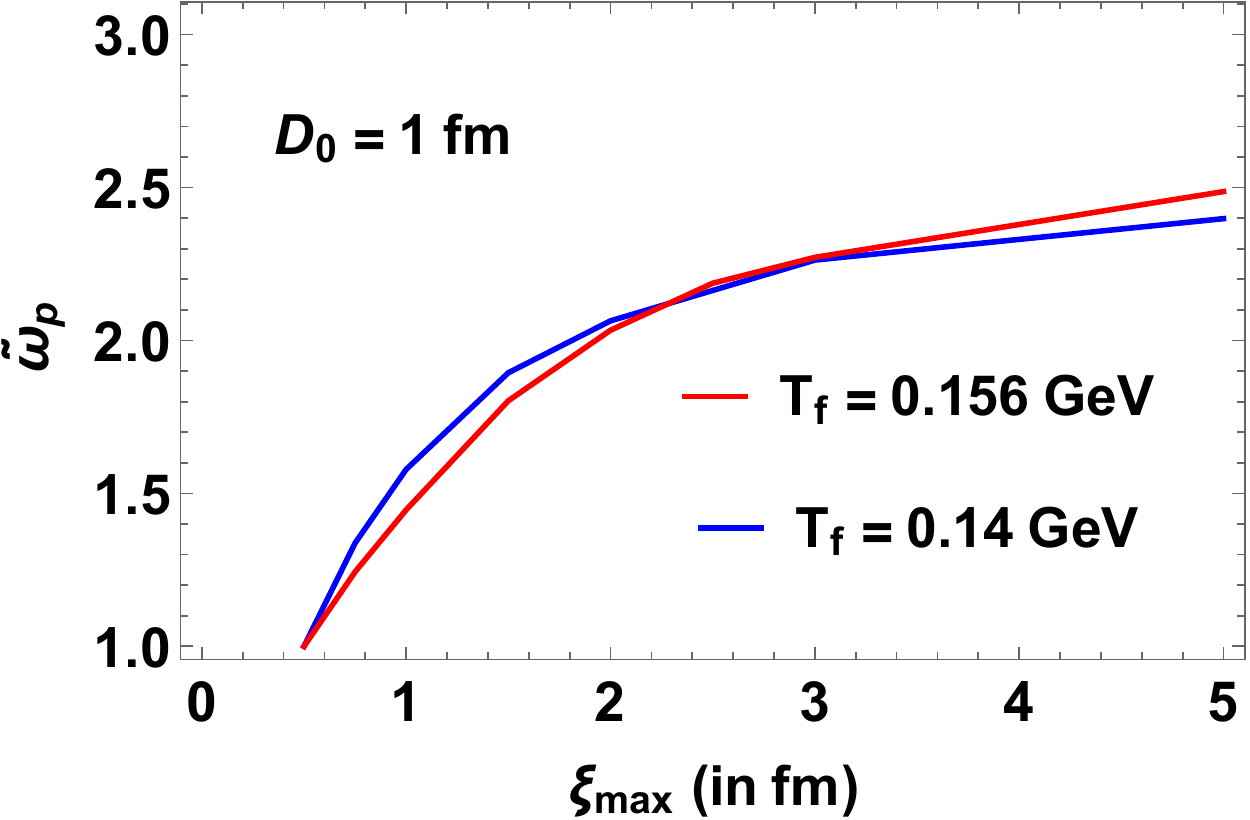}
\end{subfigure}
\begin{subfigure}{0.245\textwidth}
\includegraphics[scale=0.295]{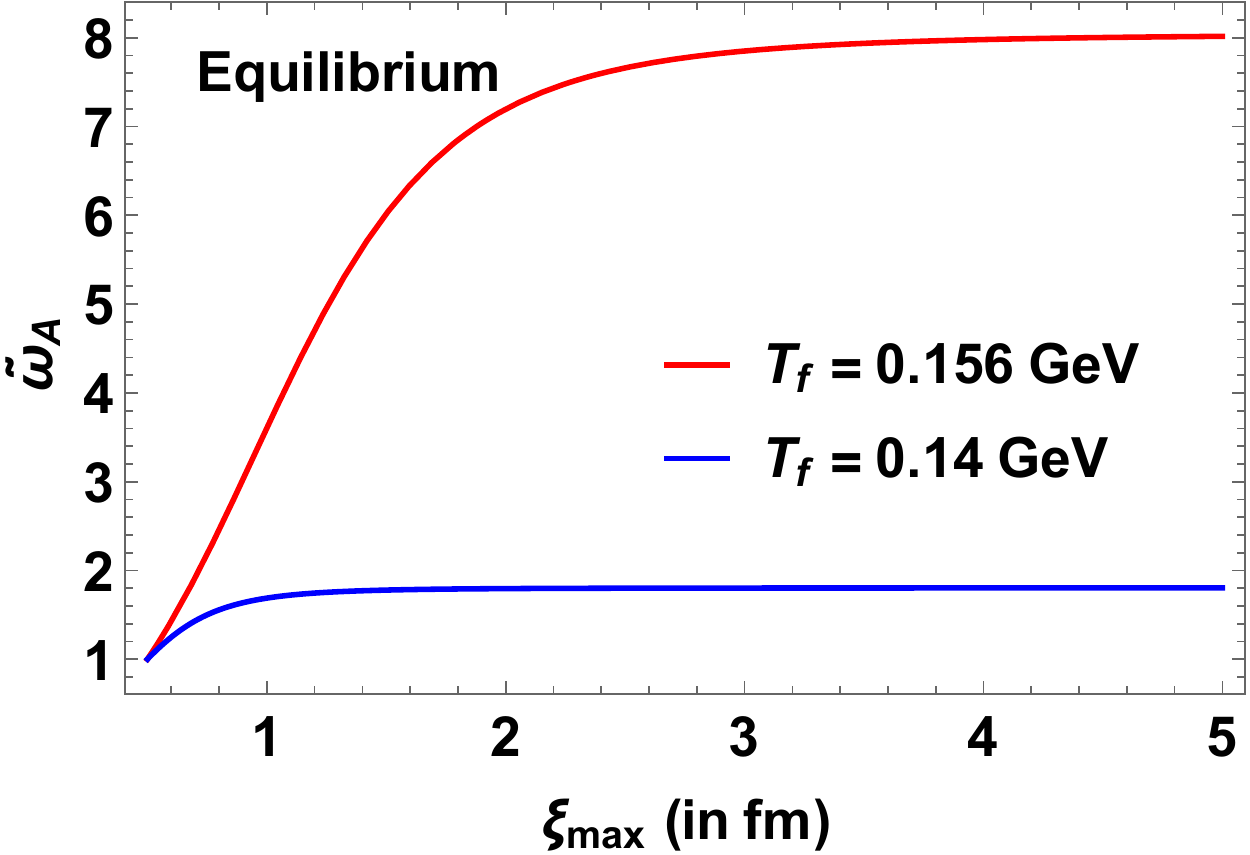}
\end{subfigure}
\end{center}
\caption{Normalized fluctuation measure $\tilde{\omega}_A$ (defined in Eq.~(\ref{omega1hydroplosfreezeout})) as a function of $\xi_{\text{max}}$ which is the maximum equilibrium correlation length reached during the hydrodynamic evolution. As $\dip\rightarrow\infty$, the $\tilde{\omega}_A$'s approach their equilibrium values .}
\label{tildeomegap}
\end{figure}

We denote the ratio of the variance defined in Eq.~(\ref{variance2}) to the mean multiplicity defined in Eq.~(\ref{mean1}) by $\omega_A$:
\begin{eqnarray}\label{omegahidyoplusfreezeout0}
\omega_A\equiv \frac{\left<\delta 
N_A^2\right>_\sigma}{\left<N_A\right>}\ .
\end{eqnarray}
The excess of the critical fluctuations over the non-critical baseline
$\omega_A^{\text{nc}}$ can be quantified via the ratio
\begin{eqnarray}\label{omega1hydroplosfreezeout}
\tilde{\omega}_A\equiv \frac{\omega_A}{\omega_A^{\text{nc}}}\ ,
\end{eqnarray}
where $\omega_A^{\text{nc}}$ is our estimate for  the value of $\omega_A$ when the correlation length is not enhanced due to proximity to a critical point, namely when it is equal to the non-critical value $\xi_0$. $\tilde{\omega}_A$ for protons  obtained within the Hydro+ framework for the simulation from Ref.~\cite{Rajagopal:2019xwg}
is shown in Fig.~\ref{tildeomegap}.

\section{Discussions and Outlook}
	
	In this work, we have introduced an extended Cooper-Frye
        procedure to freeze out critical fluctuations. We have also
        implemented it in freezing out a simplified Hydro+ simulation
        in an azimuthally symmetric and boost invariant background. We
        observe that due to critical slowing down, the
        out-of-equilibrium fluctuations did not grow as large near the
        critical point as they would have if they were able to stay in
        equilibrium. Helpfully, though, and again due to critical
        slowing down, we find the fluctuations to be less sensitive to how much lower the freeze-out temperature is than the temperature of the critical point itself than would be the case in equilibrium. This can be seen by comparing the first three panels from the left in Fig.~(\ref{tildeomegap}) to the right-most panel.
	
	The estimates that we have made in this exploratory study have relied upon many simplifications. They can be made more quantitative by employing more realistic scenarios with more appropriate initial state dynamics, a more realistic equation of state, effects of baryon stopping, and a hadronic afterburner. The coupling constant $g_A$, which sets the magnitude of the fluctuations, is not well known. However, one could constrain it by comparing the equation of state of QCD to the equation of state of a hadron resonance gas model. We have also ignored the sub-leading singularities near the critical point and the background terms in the evolution equation for the two-point correlation function. Relaxing these assumptions would be another way to improve the estimates we have presented here. The procedure also remains to be applied to obtain correlations of particles other than protons and pions in the hadron resonance gas model. 
	
	All that said, we see it as of paramount importance to extend
        the application of this procedure to higher-order non-Gauusian cumulants, as they serve as stronger signals of the critical point~\cite{Stephanov:2008qz,Athanasiou:2010kw}. While this will be the next immediate step to do, we also hope that the procedure we have developed and presented here can already be implemented within the full paradigm for the numerical simulation of  beam energy scan at RHIC~\cite{An:2021wof}.  

\section{Acknowledgement}

We gratefully acknowledge the contributions of Ryan Weller, who collaborated with us on this research project during its early stages. This work was supported by the U.S. Department of Energy, Office of Science, Office of
Nuclear Physics, within the framework of the Beam Energy Scan Theory (BEST) Topical
Collaboration and grants DE-SC0011090 and DE-FG0201ER41195. Y.Y. acknowledges support from the Strategic Priority Research Program of the Chinese Academy of Sciences, Grant No. XDB34000000.

\end{document}